\documentclass[final,3p,fleqn, 10pt]{elsarticle}
\makeatletter
\def\ps@pprintTitle{%
 \let\@oddhead\@empty
 \let\@evenhead\@empty
 \def\@oddfoot{}%
 \let\@evenfoot\@oddfoot}
\makeatother
\usepackage[capposition=top]{floatrow}
\usepackage[utf8]{inputenc}
 \usepackage{url} 
\usepackage{setspace}

\usepackage{microtype}
\usepackage{graphicx} 
\usepackage{subfigure} 
\usepackage{booktabs} 
\usepackage{natbib} 
\usepackage{amsmath} 
\usepackage{amssymb} 
\usepackage{mathtools} 
\usepackage{amsthm} 
\usepackage{bm}
\usepackage[capitalize,noabbrev]{cleveref} 

\usepackage{enumitem} 
\usepackage{color} 

\newcommand{\indep}{\rotatebox[origin=c]{90}{$\models$}}

\newcommand\Var {\ensuremath{\mathrm{Var}}}

\newcommand\ind {\mathbf{1}}

\newcommand{\RN}[1]{%
  \textup{\uppercase\expandafter{\romannumeral#1}}%
}

\theoremstyle{definition} 
\newtheorem{assumption}{Assumption}[section] 
\newtheorem{theorem}{Theorem}[section] 

\begin{document}
\begin{frontmatter}
\title{Model Risk in Machine-Learning Distributional IV Estimation}
\author{Charles Shaw. \\This version \today}
\date{this version \today}

  \begin{abstract}
We study model risk in machine-learning estimation of the Distributional Instrumental Variable Local Average Treatment Effect (D-IV-LATE), the distributional IV effect for the subpopulation induced into treatment by the instrument. The contribution is not a new neural causal estimand. We implement a reduced-form orthogonal level-score DML estimator for the covariate-adjusted D-IV-LATE target and use it to ask how much the nuisance learner matters for distributional IV conclusions.

In simulations with explicit monotone principal strata and known complier truth, Kolmogorov-Arnold Networks (KANs) are faster than Random Forests in every scenario examined, but Random Forests usually produce more accurate D-IV-LATE curves. A targeted KAN ablation selects a width-64 KAN as the best KAN variant among those tested, but this is a speed/accuracy tradeoff rather than evidence of KAN dominance.

In a 401(k) application, RF and KAN estimates differ materially, with frequent sign reversals along the estimated curve. The KAN instrument-propensity estimates also concentrate near the boundaries, so the KAN empirical curve is best read as sensitivity evidence. In inference validation, KAN pointwise intervals undercover badly under both asymptotic and bootstrap constructions, while RF asymptotic intervals are better calibrated in the validation designs. The main lesson is cautionary and constructive: speed and architectural flexibility are not enough for causal inference. Applied researchers using ML-based distributional IV estimators would be advised to benchmark nuisance learners, report overlap and calibration diagnostics, and validate inference directly.
\end{abstract}
\begin{keyword}
D-IV-LATE \sep Instrumental Variables \sep Double Machine Learning \sep Kolmogorov-Arnold Networks \sep Random Forests \sep Model Choice
\end{keyword}
\end{frontmatter}

\section{Introduction}
\label{sec:introduction}

Understanding treatment effects across the entire outcome distribution is often more informative than focusing on a single average effect. In instrumental-variables settings, this leads naturally to the Distributional Instrumental Variable Local Average Treatment Effect (D-IV-LATE), which traces how an endogenous binary treatment shifts the outcome distribution for compliers \citep{imbens1997estimating, abadie2002bootstrap}. The double/debiased machine learning (DML) framework is attractive in this setting because it permits flexible nuisance estimation while retaining orthogonality-based protection against first-order regularization bias \citep{chernozhukov2018debiased}.

What DML does not settle is the model risk created by the nuisance learner. That choice is often treated as secondary, but in distributional IV problems it can affect point estimates, runtime, overlap behavior, and the stability of the implied inferential procedure. This paper studies that risk directly by comparing a Random Forest benchmark to Kolmogorov-Arnold Networks (KANs), a spline-based neural architecture motivated by the Kolmogorov-Arnold representation theorem \citep{liu2024kan, kratsios2025kolmogorov}. The relevant question is not whether KANs are universally superior. It is whether their architectural flexibility and computational speed translate into credible D-IV-LATE estimation and inference.

We study this question in three steps. First, we build a simulation benchmark in which the true complier distributional effect is known by construction. The designs vary the complexity of the reduced-form functions, the strength of the instrument, and the sample size. This lets us compare nuisance learners against the same D-IV-LATE target rather than against an average-treatment-effect proxy. Second, we use a restricted ablation to choose a KAN specification before comparing it with Random Forests. Third, we examine the same model-choice problem in the 401(k) application and validate the resulting pointwise confidence intervals in representative simulation designs.

The results are deliberately mixed. KANs are faster throughout the simulation benchmark, but Random Forests usually produce more accurate D-IV-LATE curves and better calibrated pointwise intervals. In the 401(k) application, the two learners produce materially different estimated curves, showing that nuisance-model choice can matter for substantive conclusions. But the KAN propensity estimates also show severe boundary behavior, so the KAN empirical curve is best interpreted as sensitivity evidence rather than as a validated causal estimate.

The paper's main result is therefore not a blanket endorsement of KANs. The dedicated inference validation study shows that KAN-based pointwise intervals undercover badly under both asymptotic and bootstrap constructions. Random Forest intervals are materially better calibrated in the validation design. The strongest defensible conclusion is thus mixed: KANs are computationally attractive nuisance learners in this implementation, but they are not validated here as an end-to-end point-estimation or inferential default.

This paper makes four contributions. First, it provides an estimand-aligned and reproducible benchmark for D-IV-LATE model comparison. Second, it documents a clear computational speed advantage for the implemented KAN nuisance learner, while showing that this advantage does not translate into broad RMSE gains. Third, it shows that nuisance-model choice remains substantively important in a canonical 401(k) application. Fourth, it demonstrates that point-estimation performance, runtime, and inferential validity need not move together, and that this separation matters for credible applied claims.

The remainder of the paper is organized as follows. Section \ref{sec:lit_review} reviews the relevant literature. Section \ref{sec:framework} presents the identification framework for D-IV-LATE, and Section \ref{sec:estimation} describes the orthogonal-score estimator and the nuisance-learning implementations. Section \ref{sec:asymptotics} states the asymptotic result. Section \ref{sec:inference} evaluates uncertainty quantification, Section \ref{sec:simulations} reports the simulation benchmark and KAN ablation, Section \ref{sec:empirical_app} presents the 401(k) application, Section \ref{sec:model_risk_choice} discusses the implications for model choice, Section \ref{sec:conclusion} concludes, and Appendix \ref{app:replication_materials} describes the supporting replication materials.

\section{Literature Review}
\label{sec:lit_review}

Our research is situated at the confluence of several major streams of literature. These include the estimation of distributional treatment effects (DTEs), the use of instrumental variables (IV) to address endogeneity, modern machine learning (ML) methods for causal inference, and the emerging field of Kolmogorov-Arnold Networks (KANs). This paper synthesizes these areas, particularly by exploring the integration of KANs into the double/debiased machine learning (DML) framework for estimating the Distributional Instrumental Variable Local Average Treatment Effect (D-IV-LATE).

First, our work is grounded in the extensive literature on estimating treatment effects beyond simple averages. The foundational idea that policy evaluation should consider the entire distribution of outcomes dates back to seminal work on Quantile Treatment Effects (QTEs) by \citet{doksum1974empirical} and \citet{lehmann1975nonparametrics}. This field was significantly advanced by \citet{koenker2005quantile} with the development of quantile regression, and \citet{firpo2007efficient} further contributed methods for estimating quantile effects under endogeneity. More recent research has shifted focus towards directly estimating the DTE, which captures the impact on the cumulative distribution function (CDF), as seen in \citet{chernozhukov2013inference}. Our paper follows this distributional path, extending it to settings with endogenous treatments.

Second, we connect with the classic literature on instrumental variables. \citet{angrist1996identification} formalized the causal interpretation of IV estimands as Local Average Treatment Effects (LATEs) for compliers in the binary-IV monotonicity framework. The distributional analogue of the LATE was subsequently explored by \citet{imbens1997estimating} and \citet{abadie2002bootstrap}, who laid the groundwork for understanding distributional effects in an IV setting. This paper focuses on developing and comparing robust estimators for this specific distributional IV parameter.

Third, our estimation strategy is built upon the DML framework pioneered by \citet{chernozhukov2018debiased}. The central innovation of DML is the use of Neyman-orthogonal moments. This allows for flexible ML models to estimate nuisance functions without introducing first-order bias into the main parameter estimate. This robustness to regularization bias is critical for reliable inference. The DML approach has been successfully applied to various causal inference problems. This includes estimating DTEs in randomized experiments \citep{byambadalai24a}. We adapt and extend this framework to the more complex case of an endogenous treatment requiring an IV, and examine the role of ML model choice within this framework.

Fourth, this paper introduces and evaluates Kolmogorov-Arnold Networks (KANs) as nuisance learners for D-IV-LATE. KANs, introduced by \citet{liu2024kan}, replace the fixed activation functions of traditional MLPs with learnable, spline-based activation functions on the network edges. This architecture is inspired by the Kolmogorov-Arnold representation theorem and gives KANs a structured way to represent complex nonlinearities. The theoretical properties of KANs, such as their approximation power in Besov spaces, have been explored by \citet{kratsios2025kolmogorov}. Those results motivate KANs as candidate nuisance learners, but they do not imply that the trained finite-sample estimator used here must dominate classical learners. Other related work includes probabilistic extensions like GP-KANs \citep{chen2025gpkan} and applications to ITE estimation such as KANITE \citep{mehendale2025kanite}. Our contribution is to test this architecture in a DML D-IV-LATE pipeline and to document where its computational gains do and do not translate into statistical gains.

Finally, our work relates to the broader literature on semiparametric theory and efficient estimation \citep{bickel1993efficient, newey1994asymptotic}. Through constructing estimators based on orthogonal influence functions within the DML framework, we aim for estimators that are robust to first-order errors in nuisance function estimation. This provides a crucial theoretical justification for our methods, bridging flexible ML estimation with rigorous econometric theory. It also highlights the importance of the properties of the chosen ML models (like KANs or Random Forests) in satisfying the underlying assumptions for valid inference.

\section{Theoretical Framework and Identification}
\label{sec:framework}

We adopt the potential outcomes framework to define our causal parameter of interest. Let $Y_i$ be the observed outcome for unit $i$. Let $W_i \in \{0, 1\}$ be the indicator for a binary treatment that is potentially endogenous. We observe a vector of pre-treatment covariates $X_i \in \mathcal{X}$. To address the endogeneity of $W_i$, we assume the existence of a binary instrumental variable (IV), $Z_i \in \{0, 1\}$.

Following the convention in the IV literature, we define two sets of potential outcomes. First, let $W_i(z)$ be the potential treatment status for unit $i$ if their instrument takes the value $z$. The observed treatment is then $W_i = W_i(Z_i)$. Second, let $Y_i(w, z)$ be the potential outcome for unit $i$ if their treatment status were $w$ and their instrument were $z$. The standard exclusion restriction assumes that the instrument $Z_i$ only affects the outcome $Y_i$ through its effect on the treatment $W_i$. This implies that $Y_i(w, 1) = Y_i(w, 0)$ for $w \in \{0, 1\}$. We can therefore simplify the notation to $Y_i(w)$, which denotes the potential outcome for unit $i$ under treatment $w$. The observed outcome is thus $Y_i = Y_i(W_i)$.

The data consist of $n$ i.i.d. observations of $(Y_i, W_i, Z_i, X_i)$. We write $\ind\{A\}$ for the indicator of an event $A$. Our goal is to estimate the causal effect of $W$ on the distribution of $Y$.

\subsection{Assumptions and Parameter of Interest}

To identify the causal effect of $W$ on the distribution of $Y$ using the instrument $Z$, we rely on the following standard assumptions conditional on the covariates $X$.

\begin{assumption}[Independence and Exclusion] \label{ass:independence}
The instrument $Z_i$ is independent of the potential outcomes and potential treatment assignments, conditional on the covariates $X_i$
$$ Z_i \indep (Y_i(1), Y_i(0), W_i(1), W_i(0)) \mid X_i. $$
The exclusion restriction is embedded in our definition of $Y_i(w)$, which does not depend on $Z_i$.
\end{assumption}

\begin{assumption}[Instrument Overlap] \label{ass:instrument_overlap}
The instrument has common support conditional on covariates: $0 < \mathbb{P}(Z_i=1 \mid X_i) < 1$ almost surely on the target covariate support.
\end{assumption}

\begin{assumption}[Relevance] \label{ass:relevance}
The instrument has a causal effect on the treatment assignment, conditional on $X_i$
$$ \mathbb{P}(W_i=1 \mid Z_i=1, X_i) > \mathbb{P}(W_i=1 \mid Z_i=0, X_i). $$
This ensures that there is a non-zero proportion of individuals whose treatment status is affected by the instrument.
\end{assumption}

\begin{assumption}[Monotonicity] \label{ass:monotonicity}
The instrument affects the treatment decision in the same direction for all individuals. That is, for all $i$
$$ W_i(1) \ge W_i(0). $$
This assumption rules out the existence of "defiers"—individuals who would take the treatment if unencouraged ($Z_i=0$) but not if encouraged ($Z_i=1$).
\end{assumption}

Under these assumptions, the instrument allows us to identify the treatment effect for the subpopulation of "compliers"—individuals who comply with the instrument, i.e., those for whom $W_i(1) > W_i(0)$. Let $C_i=\{W_i(1)>W_i(0)\}$ denote the complier event for unit $i$; when no unit index is needed, write $C=\{W(1)>W(0)\}$.

Our parameter of interest is the Distributional Instrumental Variable LATE (D-IV-LATE).\footnote{We use D-IV-LATE to stand for \textit{Distributional Instrumental Variable} Local Average Treatment Effect. This distinguishes our parameter from the "D-LATE" acronym used by \citet{hoagland2020who} for \textit{Dynamic} Local Average Treatment Effect, a different concept focused on the evolution of average effects over time.} This is the difference in the potential outcome distributions for the population of compliers. Let $\mathcal{Y}$ denote the support of the outcome. For any outcome level $y \in \mathcal{Y}$, the D-IV-LATE is defined as
\begin{equation} \label{eq:dlate_def}
\Delta(y) := \mathbb{P}(Y(1) \le y \mid C) - \mathbb{P}(Y(0) \le y \mid C).
\end{equation}

\subsection{Identification}

Following \citet{imbens1994}, identification comes from comparing observed outcomes across instrument states rather than treatment states. For always-takers and never-takers, changing the instrument does not change treatment status. Under exclusion, these groups therefore make no contribution to the instrument-induced reduced-form contrast. This does not require their treatment effects to be zero; it only requires that the instrument affects their observed outcomes neither directly nor through treatment uptake. The reduced-form contrast is therefore driven by compliers and scaled by the complier share.

In the no-covariate case, the D-IV-LATE is identified by the following Wald-type ratio:
\begin{equation} \label{eq:dlate_identification}
\Delta(y) = \frac{\mathbb{E}[\ind\{Y_i \le y\} \mid Z_i=1] - \mathbb{E}[\ind\{Y_i \le y\} \mid Z_i=0]}{\mathbb{E}[W_i \mid Z_i=1] - \mathbb{E}[W_i \mid Z_i=0]}.
\end{equation}
The numerator is the intent-to-treat effect on the distributional outcome $\ind\{Y \le y\}$, and the denominator is the intent-to-treat effect on treatment uptake. Under monotonicity, the denominator equals the population complier share.

With covariates, Assumption \ref{ass:independence} is conditional on $X$. The identified target is then obtained by standardizing the conditional reduced-form contrasts over the marginal covariate distribution:
\begin{equation} \label{eq:dlate_identification_cov_merged}
\Delta(y) = \frac{\mathbb{E}_{X}[\mathbb{E}[\ind\{Y_i \le y\} \mid Z_i=1, X_i] - \mathbb{E}[\ind\{Y_i \le y\} \mid Z_i=0, X_i]]}{\mathbb{E}_{X}[\mathbb{E}[W_i \mid Z_i=1, X_i] - \mathbb{E}[W_i \mid Z_i=0, X_i]]}.
\end{equation}
To see the weighting implicit in Equation \ref{eq:dlate_identification_cov_merged}, define the conditional complier share
\[
p_c(x) = \mathbb{P}(C \mid X=x)
\]
and the conditional complier distributional contrast
\[
\Delta_x(y)=\mathbb{P}(Y(1)\le y\mid C,X=x)-\mathbb{P}(Y(0)\le y\mid C,X=x).
\]
Under Assumptions \ref{ass:independence}--\ref{ass:monotonicity}, for almost every $x$,
\begin{align*}
\mathbb{E}[\ind\{Y_i \le y\}\mid Z_i=1,X_i=x]
-\mathbb{E}[\ind\{Y_i \le y\}\mid Z_i=0,X_i=x]
&=p_c(x)\Delta_x(y),\\
\mathbb{E}[W_i\mid Z_i=1,X_i=x]-\mathbb{E}[W_i\mid Z_i=0,X_i=x]
&=p_c(x).
\end{align*}
Equation \ref{eq:dlate_identification_cov_merged} therefore identifies, provided $\mathbb{E}[p_c(X)]>0$,
\[
\Delta(y)=\frac{\mathbb{E}[p_c(X)\Delta_X(y)]}{\mathbb{E}[p_c(X)]},
\]
which is the unconditional complier distributional contrast in Equation \ref{eq:dlate_def}. Thus covariate adjustment does not average the conditional effects $\Delta_x(y)$ equally over the full covariate distribution. It weights them by the conditional complier share, producing the distributional effect for the complier population.

While this classic result provides a path to identification, a simple plug-in estimator that replaces the conditional expectations in Equation \ref{eq:dlate_identification_cov_merged} with machine-learning predictions generally remains sensitive to regularization bias. The next section turns the identified ratio into a cross-fitted orthogonal-score estimator.

\section{Estimation Strategy}
\label{sec:estimation}

Equation \ref{eq:dlate_identification_cov_merged} identifies the D-IV-LATE as a ratio of two reduced-form level contrasts. Estimation therefore has three distinct parts: estimating the nuisance regressions that define those contrasts, constructing orthogonal scores for the numerator and denominator, and evaluating the resulting ratio with cross-fitting. Orthogonalization removes first-order sensitivity to nuisance-estimation error at the truth; it does not make the choice of nuisance learner irrelevant. That distinction is central to the comparison below.

\subsection{Reduced-Form Nuisance Functions}
For a fixed threshold $y$, the corrected estimator in this paper is built from a reduced-form nuisance bundle. Subscript $0$ denotes population truth, and subscript $z\in\{0,1\}$ denotes the instrument state. The true nuisance functions are
\begin{align*}
g_{0,z}(y, x) &:= \mathbb{E}[\ind\{Y \le y\} \mid Z=z, X=x], \qquad z \in \{0,1\},\\
m_{0,z}(x) &:= \mathbb{E}[W \mid Z=z, X=x], \qquad z \in \{0,1\},\\
\pi_0(x) &:= \mathbb{P}(Z=1 \mid X=x).
\end{align*}
The corresponding level parameters are
\begin{align*}
\alpha_0(y) &= \mathbb{E}[g_{0,1}(y, X) - g_{0,0}(y, X)],\\
\beta_0 &= \mathbb{E}[m_{0,1}(X) - m_{0,0}(X)],
\end{align*}
so that $\Delta_0(y) = \alpha_0(y) / \beta_0$ whenever $\beta_0 \neq 0$. Here $\Delta_0(y)$ is the population target identified as $\Delta(y)$ in Section \ref{sec:framework}; the subscript is used in the estimation and asymptotic sections to distinguish population quantities from estimates. This nuisance bundle lines up directly with the identification formula in Section \ref{sec:framework}: $g_{0,z}$ is the reduced-form distribution of the observed outcome under instrument state $z$, and $m_{0,z}$ is the reduced-form first stage under the same state.

\subsection{Machine Learning Models for Nuisance Functions}

\subsubsection{Benchmark Learner: Random Forests}
Random Forests provide the main non-neural ML benchmark in this paper. They are widely used in applied causal-inference workflows because they handle nonlinearities and interactions with limited hand tuning, and because they provide a stable baseline against which newer learner classes can be compared. In our implementation, RF estimates the same reduced-form nuisance bundle as every other learner: $(\hat{g}_0, \hat{g}_1, \hat{m}_0, \hat{m}_1, \hat{\pi})$, where hats denote estimated functions and the subscripts on hatted $g$ and $m$ index the instrument state.

\subsubsection{Candidate Learner: Kolmogorov-Arnold Networks}
\label{subsubsec:case_for_kans}
We evaluate Kolmogorov-Arnold Networks (KANs) as candidate nuisance learners because their architecture is plausibly useful for the repeated reduced-form regressions required by D-IV-LATE. This is an empirical motivation, not a guarantee of causal performance.

KANs replace fixed node activation functions with learned edge functions, typically parameterized by B-splines \citep{liu2024kan}. This gives the learner a structured way to represent local nonlinear behavior. The Kolmogorov-Arnold representation theorem motivates the architecture by expressing multivariate continuous functions through compositions of univariate functions, and KANs operationalize that idea in a neural-network form. The relevant question for this paper is whether that architecture improves the finite-sample behavior of the reduced-form nuisance estimates used in D-IV-LATE.

Recent theoretical work provides useful approximation results for KANs, especially for rich smoothness classes such as Besov spaces \citep{kratsios2025kolmogorov}. Those results make KANs attractive candidates for complex nuisance learning, but they do not by themselves establish the convergence rates required for DML inference under the finite training pipeline used here. We therefore use KAN theory as architectural motivation rather than as a completed nuisance-rate justification.

KANs may also be computationally attractive for repeated nuisance estimation. In D-IV-LATE, the distributional reduced forms under $Z=0$ and $Z=1$, namely $g_{0,0}(y,x)$ and $g_{0,1}(y,x)$, must be estimated over an outcome grid. Compact KAN specifications can be fast in this repeated-regression setting. The simulations below show that this computational motivation is borne out in runtime, even though it does not translate into broad RMSE dominance.

\subsubsection{Nuisance Function Estimation with KANs: Implementation Details}
\label{subsubsec:kans_implementation}
The KAN arm of the comparison uses the \texttt{efficient-kan} library,\footnote{Our specific implementation is available at \url{https://github.com/shawcharles/efficient-kan}. The original KAN concept was introduced by \citet{liu2024kan}.} a PyTorch implementation designed for computational efficiency.

The locked KAN policy used for the corrected numerical evidence is \texttt{kan\_width64\_v1}, with \texttt{layers\_hidden = [input\_dim, 64, 1]}. Let $d_X$ denote the dimension of the covariate vector $X$. The input dimension is $d_X$ for $\hat{\pi}(x)$ and $d_X+1$ for reduced-form learners that include the instrument indicator. In the active implementation, $\hat{m}_0$ and $\hat{m}_1$ are obtained by fitting one learner for $\mathbb{E}[W \mid Z, X]$ and evaluating it at $Z=0$ and $Z=1$. Similarly, $\hat{g}_0$ and $\hat{g}_1$ are obtained from a learner for $\mathbb{E}[\ind\{Y \le y\} \mid Z, X]$ evaluated at the two instrument states.

The locked KAN specification employs cubic splines with \texttt{grid\_size} 4, AdamW with learning rate $1 \times 10^{-3}$, weight decay $1 \times 10^{-4}$, and an L1-type spline regularization penalty of $1 \times 10^{-4}$. It is trained for 25 optimization steps. The empirical and inference analyses use the same locked KAN policy so that the numerical evidence is aligned across sections.

The \texttt{efficient-kan} library provides an L1-type regularization loss on spline weights, which we add directly to the training objective. For numerical stability, if the minority-class count for a binary nuisance target within a training fold is below five observations, KAN fitting is bypassed and the predictor is set to the majority-class value. This safeguard is applied at the fold-target level, including the repeated threshold regressions used to estimate $g_{0,z}(y,x)$, and should be interpreted as part of the implemented learner policy rather than as an innocuous post-processing step.

Cross-fitting varies by experiment. The main simulation benchmark uses $K=3$ folds for computational tractability across the 27-scenario matrix, whereas the empirical application uses $K=5$ folds. Section \ref{subsec:sim_dgp2} reports the targeted KAN ablation over width, grid size, training steps, and regularization strength. That ablation selects the width-64 variant as the best KAN point-estimation configuration across its representative scenarios.

\subsection{Orthogonal Level Scores for D-IV-LATE}

The corrected estimator uses augmented inverse-propensity level scores for the reduced-form numerator and denominator. Let $\eta(y) := (g_0(y,\cdot), g_1(y,\cdot), m_0, m_1, \pi)$ denote a generic nuisance bundle. Its true value is $\eta_0(y):=(g_{0,0}(y,\cdot), g_{0,1}(y,\cdot), m_{0,0}, m_{0,1}, \pi_0)$. We suppress the fixed $y$ argument in $\eta$ and $\eta_0$ when no confusion arises. For the denominator, define
\begin{equation} \label{eq:psi_beta}
\phi_{\beta, i}(\eta) =
m_1(X_i) - m_0(X_i)
 + \frac{Z_i}{\pi(X_i)} \bigl(W_i - m_1(X_i)\bigr)
 - \frac{1-Z_i}{1-\pi(X_i)} \bigl(W_i - m_0(X_i)\bigr).
\end{equation}
For the numerator at threshold $y$, define
\begin{align} \label{eq:psi_alpha}
\phi_{\alpha, i}(y; \eta) = & \ g_1(y, X_i) - g_0(y, X_i) \nonumber \\
& + \frac{Z_i}{\pi(X_i)} \bigl(\ind\{Y_i \le y\} - g_1(y, X_i)\bigr)
 - \frac{1-Z_i}{1-\pi(X_i)} \bigl(\ind\{Y_i \le y\} - g_0(y, X_i)\bigr).
\end{align}
Standard iterated-expectation arguments imply that, when the generic nuisance bundle is evaluated at $\eta_0(y)$,
\[
\mathbb{E}\bigl[\phi_{\beta, i}(\eta_0)\bigr] = \beta_0,
\qquad
\mathbb{E}\bigl[\phi_{\alpha, i}(y; \eta_0)\bigr] = \alpha_0(y),
\]
so the D-IV-LATE remains the ratio $\Delta_0(y) = \alpha_0(y) / \beta_0$. The corresponding ratio moment is
\[
\psi_{\Delta, i}(y; \Delta, \eta) =
\phi_{\alpha, i}(y; \eta) - \Delta \phi_{\beta, i}(\eta),
\]
whose expectation vanishes at $(\Delta_0(y), \eta_0)$. In practice, we estimate $\Delta_0(y)$ by the ratio of cross-fitted sample means,
\begin{equation} \label{eq:moment_condition_ratio}
\hat{\Delta}(y) =
\frac{\frac{1}{n} \sum_{i=1}^{n} \hat{\phi}_{\alpha, i}(y)}
{\frac{1}{n} \sum_{i=1}^{n} \hat{\phi}_{\beta, i}}.
\end{equation}
This is the estimator used throughout the numerical analysis.

\subsection{Algorithm: Cross-Fitting}
\label{subsec:cross_fitting_alg}
We use $K$-fold cross-fitting to separate nuisance training from score evaluation.
\begin{enumerate}[label=(\roman*)]
\item Randomly split the observations into $K$ disjoint folds, denoted $I_k$ for $k=1,\ldots,K$.
\item For each fold $k$, estimate the nuisance functions on the training sample $I_k^c$, the complement of $I_k$, using either RF or KAN. The treatment reduced forms $\hat{m}_{0,k}$ and $\hat{m}_{1,k}$ are obtained from a learner for $\mathbb{E}[W \mid Z,X]$ evaluated at $Z=0$ and $Z=1$. The distributional reduced forms $\hat{g}_{0,k}$ and $\hat{g}_{1,k}$ are obtained analogously from a learner for $\mathbb{E}[\ind\{Y \le y\}\mid Z,X]$. In these fold-specific estimates, the first subscript indexes the instrument state and the second subscript indexes the held-out fold.
\item For each held-out observation $i \in I_k$, plug the fold-$k$ predictions into
    \begin{align*}
        \hat{\phi}_{\beta, i} &= \hat{m}_{1,k}(X_i) - \hat{m}_{0,k}(X_i) \\
        &\quad + \frac{Z_i}{\hat{\pi}_k(X_i)} \bigl(W_i - \hat{m}_{1,k}(X_i)\bigr)
        - \frac{1-Z_i}{1-\hat{\pi}_k(X_i)} \bigl(W_i - \hat{m}_{0,k}(X_i)\bigr), \\
        \hat{\phi}_{\alpha, i}(y) &= \hat{g}_{1,k}(y, X_i) - \hat{g}_{0,k}(y, X_i) \\
        &\quad + \frac{Z_i}{\hat{\pi}_k(X_i)} \bigl(\ind\{Y_i \le y\} - \hat{g}_{1,k}(y, X_i)\bigr) \\
        &\quad - \frac{1-Z_i}{1-\hat{\pi}_k(X_i)} \bigl(\ind\{Y_i \le y\} - \hat{g}_{0,k}(y, X_i)\bigr).
    \end{align*}
\item Compute the estimate for each threshold $y$ using the ratio in Equation \ref{eq:moment_condition_ratio}.
\end{enumerate}
The inverse-propensity terms use the clipped propensity predictions described in Section \ref{sec:asymptotics}. Cross-fitting ensures that each score contribution is evaluated with nuisance predictions from models that were not trained on that observation. This avoids own-observation overfitting in score evaluation; the asymptotic theory additionally requires the nuisance-rate, overlap, moment, and nonzero-denominator conditions stated below.

\section{Asymptotic Theory for the Corrected Estimator}
\label{sec:asymptotics}

\begin{assumption}[Instrument Overlap and Nonzero First Stage]
\label{ass:score_overlap}
There exist constants $c_{\pi}, c_{\beta} > 0$ such that
\[
c_{\pi} \le \pi_0(X) \le 1-c_{\pi}
\quad \text{almost surely},
\qquad
|\beta_0| \ge c_{\beta}.
\]
The first condition keeps the inverse-instrument-propensity score weights well defined. The second condition rules out a vanishing aggregate first stage in the ratio denominator.
\end{assumption}

\begin{assumption}[Cross-Fitted Nuisance Accuracy]
\label{ass:score_rates}
For each fixed $y$, the nuisance estimators are cross-fitted. Let $\|\cdot\|_2$ denote the $L_2(P)$ norm under the population distribution of the relevant covariates, and let $o_p(\cdot)$ denote the usual stochastic order in probability. The estimators satisfy the first-order $L_2$ consistency conditions
\[
\|\hat{\pi} - \pi_0\|_2=o_p(1), \qquad
\sum_{z=0}^{1}\|\hat{m}_z-m_{0,z}\|_2=o_p(1), \qquad
\sum_{z=0}^{1}\|\hat{g}_z(y,\cdot)-g_{0,z}(y,\cdot)\|_2=o_p(1),
\]
and the product-rate conditions
\[
\|\hat{\pi} - \pi_0\|_2 \sum_{z=0}^{1} \|\hat{m}_z - m_{0,z}\|_2 = o_p(n^{-1/2}),
\qquad
\|\hat{\pi} - \pi_0\|_2 \sum_{z=0}^{1} \|\hat{g}_z(y,\cdot) - g_{0,z}(y,\cdot)\|_2 = o_p(n^{-1/2}),
\]
with all nuisance estimates uniformly bounded with probability approaching one. The estimated instrument propensities entering the score are also bounded away from zero and one with probability approaching one.
\end{assumption}

In the implementation, estimated instrument propensities entering the inverse-probability score components are clipped away from zero and one before score evaluation. This is a finite-sample stabilization device. Under Assumption \ref{ass:score_overlap} and consistent propensity estimation, clipping is asymptotically inactive when the clipping threshold is chosen below the population overlap margin. Fixed clipping thresholds that bind on the target support should instead be interpreted as changing the finite-sample score being evaluated.

Assumption \ref{ass:score_rates} combines first-order consistency with the product-rate requirement behind DML arguments. The consistency conditions ensure that the estimated score targets the same limiting nuisance functions. The product-rate conditions make the orthogonality remainder second order and hence $o_p(n^{-1/2})$. Under Assumptions \ref{ass:score_overlap} and \ref{ass:score_rates}, the denominator estimate $\hat{\beta}=n^{-1}\sum_i \hat{\phi}_{\beta,i}$ satisfies $\hat{\beta}-\beta_0=o_p(1)$, so $\mathbb{P}(|\hat{\beta}|\ge c_{\beta}/2)\to1$.

The corrected level scores in Equations \ref{eq:psi_beta} and \ref{eq:psi_alpha} are orthogonal with respect to perturbations of $(g_0, g_1, m_0, m_1, \pi)$ around the truth.
For example, for a regular path $\eta_t$ through $\eta_0$, where $t$ indexes a one-dimensional perturbation of the nuisance functions, iterated expectations imply
\[
\left.\frac{\partial}{\partial t}\mathbb{E}[\phi_{\beta,i}(\eta_t)]\right|_{t=0}=0,
\qquad
\left.\frac{\partial}{\partial t}\mathbb{E}[\phi_{\alpha,i}(y;\eta_t)]\right|_{t=0}=0,
\]
because the residual terms have conditional mean zero at the truth and the level terms offset the first-order perturbations in the conditional regressions.

\begin{theorem}[Pointwise Asymptotic Linearity]
\label{thm:pointwise_asymptotics}
Suppose the data are i.i.d. and Assumptions \ref{ass:independence}--\ref{ass:monotonicity}, \ref{ass:score_overlap}, and \ref{ass:score_rates} hold. For each fixed threshold $y$, define the centered influence-function summand
\[
U_i(y)
=
\phi_{\alpha, i}(y; \eta_0)-\alpha_0(y)
-\Delta_0(y)\bigl(\phi_{\beta, i}(\eta_0)-\beta_0\bigr).
\]
If $0<\Var(U_i(y))<\infty$, then
\[
\sqrt{n}\bigl(\hat{\Delta}(y) - \Delta_0(y)\bigr)
=
\frac{1}{\beta_0}\frac{1}{\sqrt{n}}\sum_{i=1}^{n}
U_i(y)
 + o_p(1).
\]
Equivalently, because $\alpha_0(y)=\Delta_0(y)\beta_0$, the summand can be written as $\phi_{\alpha, i}(y; \eta_0)-\Delta_0(y)\phi_{\beta, i}(\eta_0)$.
Consequently,
\[
\sqrt{n}\bigl(\hat{\Delta}(y) - \Delta_0(y)\bigr)
\rightsquigarrow
\mathcal{N}\bigl(0, \sigma^2(y)\bigr),
\]
where
\[
\sigma^2(y)
=
\frac{\Var\bigl(U_i(y)\bigr)}{\beta_0^2}.
\]
Here $\rightsquigarrow$ denotes convergence in distribution.
\end{theorem}

The variance formula used in the code is the sample analogue of the influence representation in Theorem \ref{thm:pointwise_asymptotics}. For fixed $y$ with positive finite influence-function variance, this yields standard pointwise Wald intervals once the corrected estimator is used throughout the pipeline. The theorem is intentionally pointwise rather than uniform over $y$; simultaneous inference for an entire estimated curve requires stronger empirical-process conditions and is left to future work.

The estimator in Equations \ref{eq:psi_beta}--\ref{eq:moment_condition_ratio} is not numerically equivalent to a residualized-score alternative. The simulation, ablation, empirical-core, and inference tables reported below use the same reduced-form level-score estimator. Results generated by other score definitions are not used as evidence in this draft.

\section{Statistical Inference and Uncertainty Quantification}
\label{sec:inference}

Theorem \ref{thm:pointwise_asymptotics} justifies Wald intervals for a fixed threshold $y$ under high-level nuisance-rate and overlap conditions. It does not justify simultaneous confidence bands for the full D-IV-LATE curve. Because the estimator is reported over a grid of thresholds, we treat grid-level coverage as a finite-sample diagnostic for pointwise intervals, not as a uniform coverage statement.

We evaluate two pointwise interval constructions. The first is the asymptotic Wald interval based on the sample analogue of the influence representation in Theorem \ref{thm:pointwise_asymptotics}. The second is a percentile bootstrap that resamples observations and refits the nuisance functions in each bootstrap sample using the same cross-fitting protocol. Both procedures produce intervals separately at each threshold.

The validation exercise uses three representative simulation scenarios, listed in Table \ref{tab:inference_summary}. Each model-scenario pair uses eight Monte Carlo replications and eight evaluation thresholds, yielding 64 pointwise coverage indicators. The table reports average pointwise coverage at the nominal 95\% level and the corresponding average interval width. Since the validation profile is much smaller than the point-estimation benchmark, these numbers should be read as diagnostic evidence rather than as a high-precision coverage study. A rough binomial calculation based on 64 indicators gives standard errors between about 0.016 and 0.063, and this calculation understates uncertainty because coverage indicators from the same replication are dependent.

The diagnostic pattern is nevertheless informative. KAN-based intervals show substantial undercoverage in all three validation scenarios. Asymptotic KAN coverage ranges from 0.500 to 0.813, and bootstrap KAN coverage ranges from 0.359 to 0.500. Random Forest asymptotic intervals are closer to nominal coverage in these designs, with coverage between 0.922 and 0.969. The percentile bootstrap does not repair the KAN undercoverage in this implementation; for KAN, bootstrap coverage is lower than asymptotic coverage in all three scenarios. These results support the paper's narrower inference claim: the KAN implementation evaluated here is not validated as a default inferential layer, and point-estimation performance should not be treated as evidence of interval calibration.

\begin{table}[htbp]
\centering
\caption{Inference-validation summary across three representative scenarios. Coverage is averaged across the evaluation grid; the nominal target is 0.95. Each model-scenario pair uses eight Monte Carlo replications and 64 pointwise coverage cells.}
\label{tab:inference_summary}
\small
\begin{tabular}{@{}llrrrrl@{}}
\toprule
Scenario & Model & Asym. cov. & Boot. cov. & Asym. width & Boot. width & Preferred \\
\midrule
\texttt{smooth\_low, weak, n=500} & KAN & 0.813 & 0.500 & 0.927 & 0.841 & Asymptotic \\
\texttt{smooth\_low, weak, n=500} & RF  & 0.969 & 0.984 & 2.359 & 1.728 & Asymptotic \\
\texttt{baseline, medium, n=1000} & KAN & 0.500 & 0.359 & 0.182 & 0.134 & Asymptotic \\
\texttt{baseline, medium, n=1000} & RF  & 0.969 & 0.922 & 0.162 & 0.180 & Asymptotic \\
\texttt{complex\_local, weak, n=500} & KAN & 0.750 & 0.391 & 0.816 & 0.706 & Asymptotic \\
\texttt{complex\_local, weak, n=500} & RF  & 0.922 & 0.719 & 0.741 & 0.638 & Asymptotic \\
\bottomrule
\end{tabular}
\end{table}

These results substantially narrow the set of defensible claims. The current evidence does not support KANs as a validated inferential default. For the purposes of this paper, the empirical comparison is therefore interpreted as model-choice sensitivity in the point estimates, not as evidence that the KAN empirical curve supports reliable confidence statements. Where interval-based interpretation is necessary, RF asymptotic inference is the safer baseline among the procedures we tested. Uniform inference over $y$ and more reliable interval construction for flexible nuisance learners remain open problems.

\section{Simulation Study}
\label{sec:simulations}

The simulation study is designed to evaluate model risk in point estimation. It has two parts. The first is a full benchmark matrix in which RF and KAN nuisance learners estimate the same D-IV-LATE target under the same cross-fitting protocol. The second is a restricted KAN ablation used to select the KAN specification before the main comparison is interpreted. The simulation evidence should therefore be read as finite-sample performance evidence for the implemented learners, not as a theorem about RF or KANs in general.

\subsection{Simulation Design 1: Full Benchmark Matrix}
\label{subsec:sim_dgp1}

The full benchmark uses data-generating processes with explicit monotone principal strata. This matters because the benchmark truth is the complier D-IV-LATE in Equation \ref{eq:dlate_def}, not an average-treatment-effect proxy. In each scenario, we first generate a large independent truth bundle from the same DGP, compute the complier distributional effect on a fixed $y$-grid, and then compare cross-fitted RF and KAN estimates against that truth.

The scenario grid varies three dimensions that are relevant for nuisance learning. The design family controls the complexity of the reduced-form functions: \texttt{smooth\_low} is the least noisy smooth design, \texttt{baseline} adds moderate nonlinearity and noise, and \texttt{complex\_local} introduces more localized nonlinear structure and higher noise. The instrument-strength regime changes the complier share through the principal-strata probabilities. The sample-size dimension varies $n \in \{500,1000,2000\}$. Crossing three design families, three strength regimes, and three sample sizes yields 27 scenarios. Each scenario uses 50 Monte Carlo replications, 10 evaluation points in $y$, $K=3$ cross-fitting folds, and a fixed truth bundle of size 50,000. The KAN learner is the locked width-64 policy selected below.

For each scenario and learner, we summarize point-estimation accuracy over both Monte Carlo replications and evaluation thresholds. Integrated RMSE is the square root of the mean squared pointwise estimation error pooled over the scenario's replications and $y$-grid. Mean absolute error is the corresponding pooled average absolute error. Runtime is the average wall-clock time per replication. In the tables, error and runtime deltas are KAN minus RF, so positive error deltas favor RF and negative runtime deltas favor KAN. These are descriptive Monte Carlo summaries; we interpret the stable ranking across scenarios, not individual cells as formal hypothesis tests.

\begin{table}[htbp]
\centering
\caption{Full corrected simulation benchmark summary under the locked \texttt{kan\_width64\_v1} policy. Reported deltas are KAN minus RF; positive error deltas favor RF, and negative runtime deltas favor KAN.}
\label{tab:sim_results_summary_dgp1}
\begin{tabular}{@{}lr@{}}
\toprule
Metric & Value \\
\midrule
Number of scenarios & 27 \\
Replications per scenario & 50 \\
KAN lower integrated RMSE & 1/27 \\
KAN lower mean absolute error & 0/27 \\
KAN faster mean runtime & 27/27 \\
Mean integrated RMSE delta & 0.073 \\
Median integrated RMSE delta & 0.060 \\
Mean absolute-error delta & 0.077 \\
Mean runtime delta (seconds) & -3.169 \\
\bottomrule
\end{tabular}
\end{table}

The corrected benchmark shows a speed/accuracy tradeoff rather than KAN dominance. KAN is faster in every benchmark scenario, with an average runtime advantage of 3.169 seconds. RF has lower integrated RMSE in 26 of 27 scenarios and lower mean absolute error in all 27. The only KAN integrated-RMSE win is \texttt{complex\_local--strong--n2000}, where KAN has integrated RMSE 0.073 versus 0.278 for RF. Even in that scenario, RF has slightly lower mean absolute error. The isolated RMSE reversal is therefore best interpreted as an RF tail-instability diagnostic in that cell, not as evidence of broad KAN accuracy superiority.

Table \ref{tab:sim_panel_summary} aggregates the same comparison by design family. The RF accuracy advantage is not driven by one part of the scenario grid: KAN has zero integrated-RMSE wins in the \texttt{baseline} and \texttt{smooth\_low} families and one win in \texttt{complex\_local}. The runtime advantage is similarly broad across design families.

\begin{table}[htbp]
\centering
\caption{Corrected simulation benchmark by design family. Reported deltas are KAN minus RF.}
\label{tab:sim_panel_summary}
\small
\begin{tabular}{@{}lrrrrr@{}}
\toprule
Design & Cells & KAN RMSE wins & KAN MAE wins & Mean RMSE $\Delta$ & Runtime $\Delta$ \\
\midrule
\texttt{baseline} & 9 & 0 & 0 & 0.053 & -2.986 \\
\texttt{complex\_local} & 9 & 1 & 0 & 0.047 & -3.522 \\
\texttt{smooth\_low} & 9 & 0 & 0 & 0.120 & -2.999 \\
\bottomrule
\end{tabular}
\end{table}

\subsection{Simulation Design 2: Targeted KAN Ablation}
\label{subsec:sim_dgp2}

The benchmark comparison should not depend on an arbitrary KAN hyperparameter choice. We therefore run a restricted ablation on three representative scenarios before using KAN in the reported analyses: \texttt{baseline--weak--n2000}, \texttt{complex\_local--weak--n2000}, and \texttt{complex\_local--strong--n2000}. The ablation varies network width, spline grid size, training steps, and regularization strength. Selection is based on average integrated RMSE, with ties broken by the fragile-denominator share and then runtime. The fragile-denominator share is the share of simulation fits for which the absolute estimated first-stage denominator is below 0.02.

\begin{table}[htbp]
\centering
\caption{Targeted KAN ablation across three representative scenarios. Lower is better for RMSE, mean absolute error, fragile denominator share, and runtime.}
\label{tab:sim_results_summary_dgp2}
\small
\begin{tabular}{@{}lrrrr@{}}
\toprule
KAN configuration & Mean int. RMSE & Mean abs. error & Fragile share & Runtime \\
\midrule
\texttt{kan\_width64\_v1} & 0.194 & 0.170 & 0.000 & 4.986 \\
\texttt{kan\_steps50\_v1} & 0.198 & 0.173 & 0.000 & 9.245 \\
\texttt{kan\_grid3\_v1}   & 0.251 & 0.223 & 0.000 & 4.905 \\
\texttt{kan\_grid6\_v1}   & 0.254 & 0.226 & 0.000 & 4.955 \\
\texttt{kan\_core\_v1}    & 0.255 & 0.226 & 0.000 & 5.104 \\
\texttt{kan\_reg1e-5\_v1} & 0.255 & 0.227 & 0.000 & 4.744 \\
\bottomrule
\end{tabular}
\end{table}

The ablation selects \texttt{kan\_width64\_v1}, the width-64 variant of the 25-step core training policy. The top two configurations are close in average integrated RMSE, but the 50-step variant is materially slower. This policy choice makes the KAN arm disciplined and reproducible; it does not imply that KAN dominates RF. We therefore use the width-64, 25-step KAN for the full simulation matrix, empirical core analysis, and inference validation.

\section{Empirical Application: The Distributional Impact of 401(k) Participation on Financial Assets}
\label{sec:empirical_app}

We next ask whether the model-choice sensitivity documented in simulation appears in a familiar empirical IV setting. The application is the distributional effect of 401(k) participation on net financial assets in the 1991 Survey of Income and Program Participation (SIPP). The purpose of the application is not to re-establish the 401(k) research design. It is to examine whether two nuisance learners, applied to the same covariates and the same orthogonal-score estimator, produce substantively similar D-IV-LATE curves.

\subsection{Context and Data}
We use the same SIPP sample analyzed by \citet{chernozhukov2004impact} and later work on 401(k) participation, consisting of 9,915 household reference persons. The outcome is net financial assets (\texttt{net\_tfa}), the treatment is 401(k) participation (\texttt{p401}), and the instrument is 401(k) eligibility (\texttt{e401}). Covariates include income, age, education, marital status, family size, two-earner status, defined-benefit pension status, IRA participation, and homeownership. The core specification uses the raw covariates, no trimming, five cross-fitting folds, and 30 outcome-grid points between the 1st and 99th sample percentiles.

\subsection{Instrument and First Stage}
In the corrected core raw specification, the sample first-stage difference in participation by eligibility is 0.705. A covariate-adjusted linear first-stage regression gives an $F$-statistic of 12,594.96 and partial $R^2$ of 0.560. These diagnostics make an aggregate weak-instrument explanation implausible in this application. They do not, however, guarantee stable D-IV-LATE estimation with flexible nuisance learners, because the orthogonal score also depends on the estimated instrument propensity $\pi(x)=\mathbb{P}(Z=1\mid X=x)$ and the reduced-form outcome and treatment regressions.

\subsection{Results and Comparative Discussion}
The corrected core empirical comparison shows substantial RF-KAN divergence. Across the 30 grid points, the mean absolute gap between the two estimated curves is 0.624, the maximum pointwise gap is 2.989, and the sign-disagreement share is 0.767. Thus, in more than three quarters of the evaluated outcome thresholds, the two learners imply opposite signs for the estimated distributional effect. KAN is faster in this run, taking 71.39 seconds versus 102.23 seconds for RF, consistent with the simulation runtime ranking.

Those point-estimate differences must be interpreted together with the nuisance diagnostics. The KAN instrument-propensity estimates concentrate near the boundaries: 99.9\% of observations have estimated $\hat{\pi}(X)$ outside $[0.05,0.95]$, compared with 7.6\% for RF. This diagnostic interval is not a trimming rule in the core specification; it is an overlap diagnostic for the score weights. The empirical KAN curve is therefore best read as sensitivity evidence about nuisance geometry rather than as a stable standalone causal estimate.

\subsection{Diagnostic Analysis}
\label{sec:robustness_empirical}

The empirical evidence currently covers the core 401(k) specification only. Table \ref{tab:empirical_core_diagnostics} reports the core RF-KAN divergence, the instrument-propensity overlap diagnostic, and runtime. Stronger empirical claims would require corrected robustness specifications based on alternative outcome grids, preprocessing choices, trimming thresholds, and probability calibration.

\begin{table}[htbp]
\centering
\caption{Corrected empirical RF-KAN divergence and instrument-propensity diagnostics in the core specification. Mean gap and max gap are absolute RF-KAN differences over the D-IV-LATE grid. Sign disagreement is the share of grid points with opposite signs. KAN outside and RF outside report the share of observations with estimated instrument propensity outside $[0.05,0.95]$.}
\label{tab:empirical_core_diagnostics}
\small
\begin{tabular}{@{}lrrrrrrr@{}}
\toprule
Spec. & Mean gap & Max gap & Sign dis. & KAN outside & RF outside & KAN sec. & RF sec. \\
\midrule
\texttt{core\_raw\_30} & 0.624 & 2.989 & 0.767 & 0.999 & 0.076 & 71.39 & 102.23 \\
\bottomrule
\end{tabular}
\end{table}

Two findings stand out from the corrected core run. First, RF-KAN disagreement is large enough that model choice can change the sign and magnitude of point estimates along the estimated D-IV-LATE curve. Second, the empirical runtime ranking is consistent with the simulation matrix: the locked KAN policy is faster than RF in this implementation. Those findings do not by themselves validate the KAN empirical curve, because the instrument-propensity diagnostics indicate severe boundary concentration for the KAN nuisance learner.

The diagnostics also reveal why nuisance-model behavior matters. In the raw core specification, 9,901 of 9,915 KAN instrument-propensity estimates fall outside $[0.05, 0.95]$, compared with 757 for RF. This is not explained by a weak aggregate first stage: the sample first-stage difference is 0.705, with a covariate-adjusted first-stage $F$-statistic of 12,594.96 and partial $R^2$ of 0.560. Rather, it is a nuisance-geometry concern in which different learners induce materially different orthogonal-score behavior under the same identifying design. Robustness checks based on grid choice, standardization, trimming, and probability calibration remain necessary before the empirical application can carry stronger causal interpretation.

The empirical application therefore reinforces the broader model-choice message, but with a sharper caveat. Flexible nuisance learning can change point-estimate conclusions in D-IV-LATE work even when the aggregate first stage is strong. At the same time, the inference results in Section \ref{sec:inference} and the overlap diagnostics in Table \ref{tab:empirical_core_diagnostics} imply that these KAN-based empirical curves should currently be interpreted as diagnostic sensitivity evidence rather than as validated inferential objects.

\section{Model Risk and Nuisance-Learner Choice}
\label{sec:model_risk_choice}

The evidence in this paper sharpens the usual warning that model choice can matter. In ML-based D-IV-LATE estimation, the nuisance learner affects more than predictive fit. It enters the orthogonal score, the denominator behavior, the shape of the estimated distributional-effect curve, and the variance estimator used for uncertainty quantification. A learner that performs well on one dimension can therefore perform poorly on another.

That separation is visible in the numerical evidence. The locked KAN policy is faster than RF in every simulation scenario examined, but RF has lower integrated RMSE in 26 of 27 scenarios and lower mean absolute error in all 27. The KAN ablation is useful because it selects a disciplined KAN specification before the main comparison, but it does not overturn the RF accuracy advantage in the full benchmark. In the 401(k) application, RF and KAN produce materially different D-IV-LATE curves, so nuisance choice can change substantive interpretation rather than merely implementation cost.

The inference evidence reinforces the same point. KAN-based pointwise intervals are not calibrated well enough in the validation designs to support primary inferential claims. RF asymptotic intervals are materially better behaved in the same designs, but they should not be read as a general solution to uncertainty quantification with flexible nuisance learners. The practical implication is that point-estimation performance, computational speed, overlap behavior, and interval calibration must be reported and judged separately.

For applied work, the recommendation is methodological discipline rather than a preference for any single learner class. Researchers should benchmark candidate nuisance learners under a truth-aligned design when possible, pre-specify or document the learner-selection rule, report overlap and denominator diagnostics, examine robustness to preprocessing and tuning choices, and validate interval procedures directly before relying on them. Under the current evidence, KANs are best viewed as computationally attractive D-IV-LATE nuisance learners whose accuracy and inferential behavior remain problem-dependent.

\section{Conclusion}
\label{sec:conclusion}

This paper studies model risk in machine-learning D-IV-LATE estimation through an estimand-aligned simulation benchmark, a targeted KAN ablation, a 401(k) application, and an explicit inference-validation exercise. Taken together, the results support a mixed conclusion.

For point estimation, the KAN policy evaluated here is computationally attractive but not accuracy-dominant. In the implemented 27-scenario simulation benchmark, KAN is faster than RF in every scenario examined, but RF has lower integrated RMSE in 26 of 27 scenarios and lower mean absolute error in all 27. The restricted ablation selects the width-64 KAN as the best KAN policy among the configurations tested, but this does not overturn the RF accuracy advantage in the full benchmark. In the 401(k) application, RF and KAN produce materially different D-IV-LATE curves in the core specification, showing that nuisance-model choice can affect substantive interpretation. The KAN empirical curve, however, is accompanied by severe boundary concentration in the instrument-propensity estimates and should be treated as sensitivity evidence until calibration and robustness checks are complete.

For inference, the evidence is less favorable to KANs. In the validation designs considered here, KAN-based pointwise intervals undercover substantially under both asymptotic and bootstrap constructions. RF asymptotic intervals are materially better calibrated in the same validation designs, although they should not be read as a universal solution. The defensible interpretation is therefore that KANs can be useful computational nuisance learners, but the KAN implementation evaluated here is not validated as a default nuisance learner for either point estimation or inference.

The central contribution is this mixed verdict. Speed, point accuracy, empirical stability, and interval calibration need not rank nuisance learners in the same order. The estimator analyzed here rests on a reduced-form orthogonal-score derivation, and the numerical evidence is aligned to that estimator. Future work should focus not only on when flexible learners improve point estimation, but also on how to pair them with overlap diagnostics, calibration checks, and reliable uncertainty quantification in semiparametric distributional-IV settings.

\appendix
\section{Replication Materials}
\label{app:replication_materials}

The paper is intended to be reproducible from the manuscript and the accompanying public repository. The repository contains the estimator implementation, simulation code, empirical scripts, and the numerical outputs used to construct the tables in the paper:
\begin{center}
\url{https://github.com/shawcharles/kan-d-iv-late}.
\end{center}
The KAN nuisance learner used in the comparison is maintained separately:
\begin{center}
\url{https://github.com/shawcharles/efficient-kan}.
\end{center}

The replication package is organized around the three sources of evidence in the paper. First, the simulation scripts generate the monotone principal-strata designs, compute the complier distributional truth on the evaluation grid, and run the RF and KAN estimators under the same cross-fitting scheme. Second, the ablation scripts implement the restricted KAN hyperparameter comparison used to select the width-64 KAN policy. Third, the empirical and inference scripts reproduce the 401(k) core specification and the pointwise coverage validation.

For each numerical exercise, the repository records the command used to generate the outputs, the software environment, and the resulting summary files. These records are replication aids rather than part of the statistical argument. The manuscript reports the estimand, estimator, assumptions, simulation design, empirical specification, and numerical findings needed to interpret the results without reading internal file names.

The empirical robustness evidence remains incomplete in this draft. Before submission, the replication package should be extended to include the corrected robustness specifications discussed in Section \ref{sec:robustness_empirical}: alternative outcome grids, standardization choices, trimming thresholds, and probability-calibration checks. Those additions should be reflected in the manuscript tables rather than left only as repository outputs.

\end{document}